\documentclass[twocolumn,prd,aps,floatfix,showpacs]{revtex4}
\usepackage{graphicx}

\def\Mp{M_{\ast}}
\def\beqra{\begin{eqnarray}}
\def\eeqra{\end{eqnarray}}
\def\beq{\begin{equation}}
\def\eeq{\end{equation}}
\def\ds{\displaystyle}

\def\vp{\varphi}

\begin{document}
\input epsf


\title{Dark Matter Relic Abundance and Scalar-Tensor Dark Energy}
\author{R.~Catena$^{(1)}$, N.~Fornengo$^{(2)}$, A.~Masiero$^{(3,4)}$, M.~Pietroni$^{(4)}$ and F.~Rosati$^{(3,4)}$} 
\address{$^{(1)}${\it Scuola Normale Superiore and INFN - Sezione di Pisa, 
Piazza dei Cavalieri 7, I-56125 Pisa, Italy  }}
\address{$^{(2)}${\it Dipartimento di Fisica Teorica, Universit\`a di Torino and \\ INFN - Sezione di Torino, 
via P. Giuria 1 I-10125 Torino, Italy  }}
\address{$^{(3)}${\it Dipartimento di Fisica, Universit\`a di Padova,  via Marzolo 8, I-35131, Padova, Italy }}
\address{$^{(4)}${\it INFN, Sezione di Padova, via Marzolo 8, I-35131, Padova, Italy}}
\address{\rm e--mail: r.catena@sns.it, fornengo@to.infn.it,
masiero@pd.infn.it, pietroni@pd.infn.it, rosati@pd.infn.it}

\begin{abstract}
Scalar-tensor theories of gravity provide a consistent framework to
accommodate an ultra-light quintessence scalar field. While the
equivalence principle is respected by construction, deviations from
General Relativity and standard cosmology may show up at
nucleosynthesis, CMB, and solar system tests of gravity. After
imposing all the bounds coming from these observations, we consider
the expansion rate of the universe at WIMP decoupling, showing that it
can lead to an enhancement of the dark matter relic density up to few
orders of magnitude with respect to the standard case. This effect can
have an impact on supersymmetric candidates for dark matter.
\end{abstract}
\pacs{98.80.Cq, 98.80.-k, 95.35.+d \hspace{3 cm}  DFTT 10/2004, \,  DFPD/04/TH/08}

\maketitle

%
%
\section{Introduction}
%
According to our current understanding \cite{rev-cosmo}, Dark Matter
(DM) and Dark Energy (DE) represent the two major components of the
present Universe.  Surprisingly, it is found that the DM and DE energy
densities, $\rho_{DM}$ and $\rho_{DE}$, are today roughly the same
(differing only by a factor of two), while their ratio has been
varying by several orders of magnitude in the past history of the
Universe.

It seems quite natural, then, to explore the possibility of a DM--DE
interaction which could account for this coincidence. This approach, however,
is not free from problems if the DE component is interpreted in
terms of a dynamical quintessence \cite{quint-review} scalar field.
Indeed, such a scalar is constrained to be extremely light in order to
fit the data, giving rise to unwanted long-range forces which may
represent a severe threat to the equivalence principle. In addition,
couplings of the quintessence scalar with the gauge field strengths
are potential sources of dangerous time variations of the fundamental
constants\footnote{See, for example,
Refs. \cite{carroll1,mpr,dpv,dam3a,dam3b} for different approaches on these
problems.}.

A possible way of coping with a very light scalar while avoiding these
shortcomings, is to choose to work in the framework of a scalar-tensor
gravity (ST) theory \cite{bd}, where by construction matter has a
purely metric coupling with gravity.  It has been shown \cite{max}
that in this case, the scalar field can benefit from an attraction
mechanism which, during the matter dominated era, makes ST overlapping
with standard General Relativity (GR). At the same time, ST may
possess a second ``attraction mechanism'' \cite{max} which will ensure
the correct evolution of $\rho_{DE}$ along a so-called `tracking'
\cite{tracker} solution.

While ST can very closely reproduce standard GR at the present time,
it may however lead to major differences in the past evolution of the
Universe, differences which may result in observable consequences for
us today.  For example, it has been shown
\cite{joyce,damour-pichon,santiago,carroll2} that ST theories may have
a profund impact on nucleosynthesis.  At the same time, a curious fact
has recently come to attention: a non--conventional dynamics of the
quintessence scalar in the past history of the universe may remain
`hidden' to the available cosmological observations, but manifest
itself through the DM relic abundance
\cite{salatirosati,ullio,comelli}.  It is then worth studying if ST
may provide a viable quintessence candidate and at the same time have
an impact on the DM relic abundance. We have considered this
possibility and computed explicitly the differences from the standard
case.

When considering ST theories, the departures from standard cosmology
are mainly due to the different expansion rate ($\tilde{H}$) which they determine.
Such deviation from the usual expansion rate of GR bears potentially
relevant consequences
in all those phenomena which are closely dependent on the
timing in which they occur.  The aim of the present work is to study the
possible modifications of the expansion rate of the Universe in ST at
the time of Cold Dark Matter (CDM) freeze-out, focussing on Weekly
Interacting Massive Particles (WIMPs) as the most natural candidates
for CDM.  As it is well known, their present relic density depends on
the precise moment they decouple and, in turn, on the precise moment
the WIMP annihilation rate equals the expansion rate of the Universe.
We expect then that a variation of $\tilde{H}$ in the past may lead to
measurable consequences on the WIMP relic abundance.
      
In order to assess the allowed departure of $\tilde{H}$ from its standard value
at WIMPs freeze-out, we have to take into account the bounds imposed
on ST by phenomena at later epochs \cite{Riaz}.  We already mentioned
the crucial test of nucleosynthesis: passing the BBN exam will imply
rather severe restrictions on the coupling of the scalar field of ST
to matter and also on its initial conditions at temperatures much
higher than the WIMP freeze-out. Coming to later epochs, we have to
consider the photon decoupling and the consequent restrictions imposed
by CMB data \cite{CMB}. 
Interestingly enough, we will show that the BBN
``filter'' on ST is so efficient that it drastically limits any visible
effect on the CMB spectrum (in particular, shifts in the peaks
positions are forced to be smaller than the experimental error). More
restrictive than the CMB probe turn out to be the GR tests, in
particular after the tight bound on the post-newtonian parameter
$\gamma$, recently provided by the Cassini spacecraft
\cite{cassini}. This constraint becomes quite relevant when combined
with that on the scalar equation of state of $w_\phi$ coming from SNe
Ia data \cite{SNe}. We will find that significant deviations from
standard cosmology are possible only if the DE equation of state
differs appreciably from $-1$ today. In other words, if DE is a
cosmological constant, it is unlikely that future cosmological
observations will be able to discriminate between ST and GR.

The question we intend to explicitly tackle is the following: taking
all the abovementioned restrictions (BBN, CMB, GR tests) into account,
how much can the Hubble parameter $\tilde{H}$ at the time of WIMP freeze-out
differ from its canonical value if ST replaces GR?  In other words,
how much is the WIMP relic density allowed to vary, if we consider ST
instead of GR?

We find that in ST theories the expansion rate of the Universe at few
GeVs can profoudly differ from the usual value obtained in GR (with
variations up to five orders of magnitude) and, yet, allow the correct
light elements production at BBN.  This situation is perfectly
analogous to the `kination' effect studied in \cite{salatirosati},
where a modification of $\tilde{H}$ at WIMP freeze-out was induced by a short
period of dominance of the scalar kinetic energy, although with some
deal of fine-tuning.  In the case considered here, the effect of ST on
$\tilde{H}$ depends on the strength of the scalar-matter coupling, however no
particular fine-tuning is needed to pass the severe nucleosynthesis
test even when large modifications of $\tilde{H}$ at freeze-out occur. This
means that the attraction of ST towards GR proceeds very rapidly
during the cooling of the Universe from the few GeVs of WIMPs
freeze-out down to the MeV range of nucleosynthesis. The overlap of ST
with GR can subsequently be very efficient leading to ST scenarios
which can hardly be disentangled from ordinary GR in present tests at
the post-newtonian level. The fact that ST strongly affects the number
of CDM particles may turn out to be the major signature of these
theories.

>From the point of view of particle physics model building, these large
variations in the WIMPs number density today is of utmost
relevance. Particles which were not considered suitable to play a
significant role in CDM scenarios can be rescued because of their
enhanced number density.  On the other hand, particles (or regions of
the parameter space for certain WIMPs candidates), which in usual GR
scenarios constitute promising CDM candidates, would be excluded
because their boosted number would overclose the Universe. These
considerations become of particular interest if we focus on the case
where the WIMPs correspond to the lightest supersymmetric particle. A
complete analysis of the cosmologically excluded and cosmologically
interesting regions of the SUSY parameter spaces in different SUSY
contexts, when ST is considered, is presently in progress
\cite{inprogress}.

%
%
\section{Scalar-tensor theories of gravity}
%
ST theories represent a natural framework in which massless scalars
may appear in the gravitational sector of the theory without being
phenomenologically dangerous. In these theories a metric coupling of
matter with the scalar field is assumed, thus ensuring the equivalence
principle and the constancy of all non-gravitational coupling
constants \cite{dam}. Moreover, as discussed in \cite{dam3a,dam3b}, a large
class of these models exhibit an attractor mechanism towards GR, that
is, the expansion of the Universe during the matter dominated era
tends to drive the scalar fields toward a state where the theory
becomes indistinguishable from GR.

ST theories of gravity are defined by the action \cite{dam,dam3a,dam3b}
\begin{equation}
S=S_{g}+S_{m}\,,  \label{jordan0}
\end{equation}
where 
\beqra
&&S_{g}=\frac{1}{16\pi }\int d^{4}x\sqrt{-\tilde{g}}\left[ \Phi^2 \tilde{R} \; + \right.\nonumber \\
&& \quad\quad+
\left. 4 \,\omega (\Phi ) \tilde{g}^{\mu \nu }\partial _{\mu }\Phi
\partial _{\nu }\Phi -4\tilde{V}(\Phi )\right] \,.  \label{jordan}
\eeqra
The matter fields $\Psi _{m}$ are coupled only to the metric tensor $%
\tilde{g}_{\mu \nu }$ and not to $\Phi $, {\it i.e.} $S_{m}=S_{m}[\Psi _{m},%
\tilde{g}_{\mu \nu }]$. $\tilde{R}$ is the Ricci scalar constructed from the
physical metric $\tilde{g}_{\mu \nu }$. Each ST model is identified by the
two functions $\omega (\Phi )$ and $\tilde{V}(\Phi )$. For instance, the
well-known Jordan-Fierz-Brans-Dicke (JFBD) theory \cite{bd} corresponds to $%
\omega (\Phi )=\omega $ (constant) and $\tilde{V}(\Phi )=0$. 

The matter energy-momentum tensor is conserved, masses and
non-gravitational couplings are time independent, and in a locally
inertial frame non gravitational physics laws take their usual
form. Thus, the `Jordan' frame variables $\tilde{g}_{\mu \nu }$ and
$\Phi $ are also denoted as the `physical' ones in the literature. On
the other hand, the equations of motion are rather cumbersome in this
frame, as they mix spin-2 and spin-0 excitations. A more convenient
formulation of the theory is obtained by defining two new
gravitational field variables, $g_{\mu \nu }$ and the dimensionless
field $\varphi $, by means of the conformal transformation
\begin{eqnarray}
\tilde{g}_{\mu \nu } &\equiv &\displaystyle A^{2}(\varphi )g_{\mu \nu } 
\nonumber \\
\Phi^2 &\equiv &\displaystyle 8 \pi M_{\ast}^2 A^{-2}(\varphi ) \nonumber \\
V(\varphi ) &\equiv &\displaystyle \frac{A^{4}(\varphi )}{4 \pi} \tilde{V}(\Phi )  \nonumber \\
\alpha (\varphi ) &\equiv &\displaystyle \,\frac{d\log A(\varphi )}{d\varphi }\,.
\label{transf}
\end{eqnarray}
Imposing the condition 
\begin{equation}
\alpha ^{2}(\varphi )=\frac{1}{4\omega (\Phi )+6}\,,
\end{equation}
the gravitational action in the `Einstein frame' reads 
\begin{equation}
S_{g}=\frac{M_{\ast}^2}{2}\int d^{4}x\sqrt{-{g}}\left[ {R}+{g}^{\mu
\nu }\partial _{\mu }\varphi \partial _{\nu }\varphi -\frac{2}{M_{\ast}^2} V(\varphi )\right] ,
\end{equation}
and matter couples to $\vp$ only through a purely metric coupling, 
\begin{equation} 
S_m = S_{m}[\Psi_{m},A^{2}(\varphi ){g}_{\mu \nu }] \,\,\, . 
\end{equation}
In this frame masses and non-gravitational coupling constants are
field-dependent, and the energy-momentum tensor of matter fields is
not conserved separately, but only when summed with the scalar field
one. On the other hand, the Einstein frame Planck mass $M_{\ast }$ is
time-independent and the field equations have the simple form
\begin{eqnarray}
&&R_{\mu \nu }-\mbox{\small $\frac{1}{2}$}g_{\mu \nu }R =\displaystyle%
\frac{T^\vp_{\mu \nu }}{M_{\ast}^2} +\frac{T_{\mu \nu }}{M_{\ast}^2}  \nonumber \\
&&  \nonumber \\
&& \displaystyle \partial ^2 \varphi +\frac{1}{\Mp^2} \frac{\partial V}{\partial \varphi } = -\frac{1}{M_{\ast}^2}\frac{\alpha (\varphi )}{\sqrt{2}}T
 \,,
\label{eomgen}
\end{eqnarray}
where 
\[
T^\vp_{\mu \nu} = M_\ast^2 \partial _{\mu }\varphi \partial _{\nu }\varphi -g_{\mu \nu }\left[M_\ast^2 \frac{g^{\rho \sigma }}{2} \partial _{\rho }\varphi \partial _{\sigma }\varphi
-V(\varphi )\right] \,,
\]
and $T_{\mu \nu } = 2(-g)^{-1/2}\, \delta S_m /\delta g^{\mu\nu}$ is
the matter energy-momentum tensor in the Einstein frame.  The relevant
point about the scalar field equation in (\ref{eomgen}) is that its
source is given by the trace of the matter energy-momentum tensor, $
T\equiv g^{\mu\nu}T_{\mu\nu}$, which implies the (weak) equivalence
principle. Moreover, when $\alpha(\varphi)=0$ the scalar field is
decoupled from ordinary matter and the ST theory is indistinguishable
from ordinary GR.

We next consider an homogeneous cosmological space-time 
\[
ds^{2}= dt^{2} - a^{2}(t)\ dl^{2}\ , 
\]
where the matter energy-momentum tensor admits the perfect-fluid representation 
\[
T^{\mu \nu }=(\rho +p)\ u^{\mu }u^{\nu }\ - p\ g^{\mu \nu }\ \ , 
\]
with $g_{\mu \nu }\ u^{\mu }u^{\nu }=1$.

The Friedmann-Robertson-Walker (FRW) equations then take the form 
\begin{eqnarray}
\frac{\ddot{a}}{a} &=& - \frac{1}{6 \Mp^2} \left[ \rho +3\ p +2 \Mp^2\dot\varphi^{2}-2V(\varphi ) \right]
\\
\left( \frac{\dot{a}}{a}\right) ^{2}+\frac{k}{a^{2}} &=& \frac{1}{3 \Mp^2}
\left[\rho + \frac{\Mp^2}{2}\dot\varphi^{2}+V(\varphi ) \right] \\
\ddot{\varphi} +3\frac{\dot{a}}{a}\dot\varphi &=& 
-\frac{1}{M_{\ast}^2}\left[ \frac{\alpha (\varphi )}{\sqrt{2}}(\rho-3p)+  \frac{\partial V}{\partial \varphi } 
 \right]
 \ ,
\end{eqnarray}
with the Bianchi identity 
\begin{equation}
{}d(\rho a^{3})+p\ da^{3}=(\rho -3\ p)\ a^{3}d\log A(\varphi ). 
\label{bianchi}
\end{equation}

The physical proper time, scale factor, energy, and pressure, are related
to their Einstein frame counterparts by the relations 
\[
d\tilde{\tau }=A(\varphi )d\tau\,, \;\;\tilde{a}=A(\varphi)a\,,
\;\;
\tilde{\rho }=A(\varphi )^{-4}\rho\,,\;\;
\tilde{p}=A(\varphi )^{-4}p. 
\]

\bigskip Defining new dimensionless variables
\[
N \equiv \log \frac{a}{a_0},\qquad \qquad 
\lambda \equiv \frac{V(\varphi )}{\rho },\qquad \qquad w \equiv \frac{p}{\rho }, 
\]
and setting $k=0$ (flat space) the field equation of motion takes the more
convenient form  
\begin{eqnarray}
 && \frac{2}{3}\;\frac{1+\lambda }{1-{\varphi ^{\prime }}^{2}/6}\ \varphi ^{\prime \prime}+[(1-w )+2\lambda ] \varphi ^{\prime}= \nonumber \\
&&\quad\quad\quad\quad\quad - \sqrt{2}\; \alpha (\varphi ) \;(1-3 w)-2 \;\lambda \;\frac{d\log V(\varphi )}{d\varphi },
\label{eom}
\end{eqnarray}
where primes denote derivation with respect to $N$.  
This will be our master equation.

The effect of the early presence of a scalar field on the physical processes will come through 
the Jordan-frame Hubble parameter $\tilde H \equiv d \log \tilde{a}/d\tilde{\tau}$:
\begin{equation}
\tilde{H} = H \, \frac{(1 + \alpha (\varphi) \, \varphi ^{\prime})}{A(\vp)} \, , 
\label{Htilde}
\end{equation}
where $H\equiv d\log a/d \tau$ is the Einstein frame Hubble parameter.
In the flat--space case ($k=0$), Eq~(\ref{Htilde}) finally gives:
\begin{equation}
\tilde{H}^2 = \frac{A^2(\vp )}{3M_*^2} \, 
\frac{\left( 1 + \alpha (\vp )\,\vp ^{\prime} \right) ^2}{1-(\vp ^{\prime})^2 /6} \, 
\left[\tilde {\rho} + \tilde {V} \right] \,\,\,\, .
\label{Htilde2}
\end{equation}

%
%
\section{Evolution of the field}

%
%
\subsection{Radiation domination}
During radiation domination the scalar field energy density is
suppressed, $\lambda \ll 1$, and the first term in the RHS of
Eq.~(\ref{eom}) is proportional to
\beqra && 1-3w= \frac{\rho_{\rm
tot}- 3 p_{\rm tot}}{\rho_{\rm tot}} = \frac{\tilde{\rho}_{\rm tot}- 3
\tilde{p}_{\rm tot}}{\tilde{\rho}_{\rm tot}}\nonumber \\
&&\quad\quad\quad\quad =\frac{1}{\tilde{\rho}_{\rm tot}} \left[\sum_A
(\tilde{\rho}_A- 3 \tilde{p}_A ) + \tilde{\rho}_m\right]\,,
\label{RHS}
\eeqra where the sum runs over all particles in thermal equilibrium,
while $\tilde{\rho}_m$ is the contribution from the decoupled and
pressureless matter abundance.

\begin{figure}[t]
\epsfxsize=3.2 in 
\epsfbox{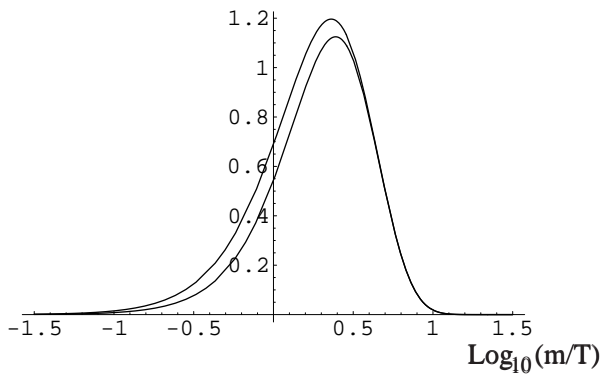}
\caption{ The function $y^2 F[y]$, with $y\equiv m/T$ defined by
Eq. (\ref{Ffm}). The upper (lower) curve corresponds to bosons
(fermions).}
\label{Fm}
\end{figure}

During radiation domination, $\tilde{\rho}_{\rm tot} \simeq \pi^2
T^4/30$, where $T$ is the Jordan-frame temperature,
 and the contribution from a single particle in equilibrium gives
\begin{equation}
\frac{\tilde{\rho}_A- 3 \tilde{p}_A}{\tilde{\rho}_{\rm tot}}\simeq\frac{15}{\pi^4}\frac{g_A}{g_\star} y_A^2  F[y_A]\,,
\end{equation}
with $y_A\equiv m_A/T$, $g_A$ the number of degrees of freedom of A, $g_\star$ the number of relativistic degrees of freedom and
\begin{equation}
F[y_A] \equiv
\int_0^\infty dx \frac{x^2}{\varepsilon_A [\exp(\varepsilon_A) \pm 1]}\,,
\label{Ffm}
\end{equation}
where $\varepsilon_A\equiv (y_A^2+x^2)^{1/2}$ and the minus (plus)
sign in the denominator of the integrand holds for bosons (fermions).
In Fig.~\ref{Fm} we plot $y^2 F[y]$. We see that it is different
from zero only around $y=1$, that is, for $T\simeq
m_A$. For higher temperatures it is quadratically suppressed
in $y$, approaching the relativistic regime in which $1-3 \tilde{w}_A
=0$. For lower temperatures it is Boltzmann- suppressed. Then, as
emphasized in \cite{dam3a,dam3b}, the field $\varphi$ evolves even during
radiation domination, receiving a `kick' each time a particle in
equilibrium becomes non-relativistic.  The second term in
Eq.~(\ref{RHS}) is suppressed as $T_{\rm eq}/T$, so it
becomes relevant only as equivalence is approached.

In the following, we will consider the evolution of the field from the
freeze out of the DM particles down to today, so we will have to take
into account all particles of masses between the freeze-out and
matter-radiation equivalence.

\subsection{Matter domination}
During matter domination $1-3w \simeq 1$ and, as long as the field
energy density is subdominant ($\lambda \ll 1 $), the RHS of the
equation of motion (\ref{eom}) is given by $-\sqrt{2} \alpha(\varphi)$
and the field evolution depends on the form of the coupling function
$\alpha(\varphi)$.  As already mentioned, the JBD theory is given by a
constant $\alpha$, and the value $\alpha=0$ corresponds to GR.  A very
attractive class of models is that in which the function
$\alpha(\varphi)$ has a zero with a positive slope, since this point,
corresponding to GR, is an attractive fixed point for the field
equation of motion \cite{dam3a,dam3b}.

It was emphasized in Ref. \cite{max} (see also \cite{sabino2}) that the fixed point starts to be
effective around matter-radiation equivalence, and that it governs the
field evolution until recent epochs, when the quintessence potential
becomes dominant. If the latter has a run-away behavior, the same
should be true for $\alpha(\varphi)$, so that the late-time behavior
converges to GR.

\subsection{Late-time beahavior}
The evolution of the field during the last redshifts depends on the
nature of DE. We will consider two possibilities: a cosmological
constant and a inverse-power law scalar potential for $\varphi$, which
can be collectively represented by the potential
\begin{equation}
V(\varphi)=\Lambda^4 \varphi^{-\delta}\qquad\qquad (\delta \ge 0),
\label{potential}
\end{equation} 
$\delta=0$ corresponding to the cosmological constant. 

In general, a cosmological constant in the Einstein frame does not
correspond to a cosmological constant in the Jordan frame, as one can
read from Eq.~(\ref{transf}). However, present tests of GR (see next
section) imply that at late times $A(\varphi) \simeq 1$, so that the
two frames are almost coincident and the expansion histories during
the last few redshifts are practically indistinguishable.

For the purpose of this paper, that is the analysis of the impact of
DE on ST cosmology, the situation in which the quintessence field is
different from $\varphi$ and decoupled from it would be basically the
same as that with a cosmological constant, since in both cases the
second term in the RHS of Eq.~(\ref{eom}) vanishes.

The main feature of the potential in (\ref{potential}) for $\delta >0$
is the existence of `tracker' solutions, which are attractors in field
space \cite{tracker}.  In the $\alpha \to 0$ limit, the late-time
behavior is completely determined by the two parameters
$\bar{\lambda}\equiv \Lambda^4/\rho_M^0$ and $\delta$.  A
non-vanishing $\alpha$ would modify the behavior of the field today,
hopefully keeping the desirable property of insensitivity to the
initial conditions.

In this paper, we will consider the following choice for $A(\varphi)$,
\beq A(\varphi)=1 +B e^{-\beta \varphi}\,,
\label{aphi}
\eeq
corresponding to 
\beq
\alpha(\varphi) = -\frac{\beta B e^{-\beta \varphi}}{1 +B e^{-\beta \varphi}}\,,
\label{alphaphi}
\eeq which has a run-away behavior with positive slope, as required by
the discussion at the end of the previous subsection. The choice for
the parameters $B$ and $\beta$ will be discussed in the following
section.

\begin{figure}[h]
\epsfxsize=3.2 in 
\epsfbox{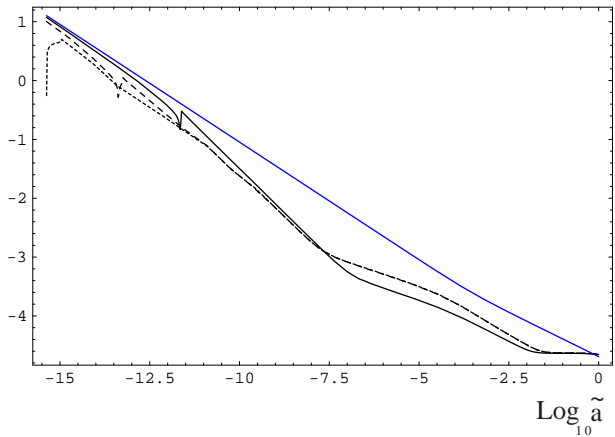}
\caption{Evolution of the energy density of the background (upper solid line)
and of three typical solutions for the scalar field. 
We see that different initial conditions converge to the same solution.}
\label{run}
\end{figure}

In Fig.~\ref{run} we show the evolution of the background and field
energy density. We see that field energy densities corresponding to
different initial conditions converge to the same solution, driven by
$\alpha(\vp)$. Notice also that the $\alpha$-attractor becomes
effective even before matter-radiation equivalence, due to the
non-relativistic decoupling effect explained above.

\begin{figure}[h]
\epsfxsize=3.2 in 
\epsfbox{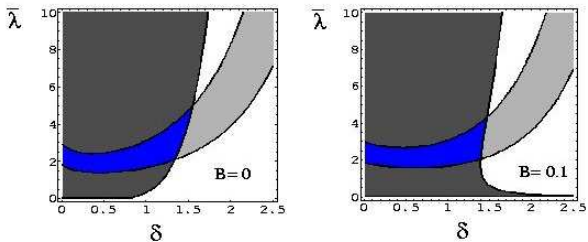}
\caption{The regions in the $\bar{\lambda}$-$\delta$ parameter plane
giving $w_\vp<-0.7$ (dark grey) and $0.65<\Omega_\varphi<0.75$ (light
grey). The left plot is the pure GR case ($B=0$) while the right one
is for ST with $B=0.1$, $\beta = 8$.}
\label{levels}
\end{figure}

In Fig.~(\ref{levels}) we show the region of parameter plane
$\bar{\lambda}$-$\delta$ giving $w_\vp<-0.7$, where
\begin{equation}
\ds
w_\varphi=\frac{\Mp^2/2 \;\dot{\varphi}^2-V(\varphi)}{\Mp^2/2\;\dot{\varphi}^2+V(\varphi)}\,,
\end{equation} 
and $0.65<\Omega_\varphi<0.75$. We see that in the ST case, ($B\neq
0$) the region giving more negative values of the equation of state is
somehow enlarged with respect to pure GR quintessence. However, in the
observationally allowed region for $\Omega_\vp$ the influence of the
$B$ parameter is negligible.

%
%
\section{Phenomenological bounds}
%
%
\subsection{Nucleosynthesis}
Assuming $\alpha \vp ^{\prime} \simeq 0$ in Eq.~(\ref{Htilde2}) -- as
we have checked numerically for the solutions relevant in this
analysis -- the Jordan frame expansion rate during nucleosynthesis may
be approximated as 

\beq 
\tilde{H}^2 
\simeq A^2(\vp) \frac{1}{3 \Mp^2}
\tilde{\rho}\,.  
\eeq 

The above expression should be compared to the
GR one, in which the Planck mass at nucleosynthesis was the same as
today. We obtain: 

\beq 
\left.\frac{\Delta\tilde{H}^2}{\tilde{H}^2}\right|_{\rm nuc}
\equiv \left. \frac{\tilde{H}^2 - \tilde{H}_{\rm GR}^2}{\tilde{H}_{\rm GR}^2}\right|_{\rm nuc}
=\frac{A^2(\vp_{nuc})-A^2(\vp_0)}{A^2(\vp_0)}\,.  
\eeq

The change in the expansion rate is completely analogous to that
obtained by adding extra neutrinos to the standard GR case. Using the
bound $\Delta N < 1$ (which is more restrictive than those obtained
for instance in \cite{nubound}), we get 
\beq
\frac{A(\vp_{nuc})}{A(\vp_0)} <1.08\,.
\label{BBN}
\eeq

\subsection{General relativity tests}
At the post-newtonian level, the deviations from GR may be
parametrized in terms of an effective field-dependent newtonian
constant
\footnote{ Strictly speaking, this is only true for a massless field,
but for any practical purpose it applies to our nearly massless
scalar ($m_\varphi\sim H_0^{-1}$) as well.}
\[
G=G(\varphi)\equiv G_* A(\varphi)^2 (1+\alpha^2(\varphi))\,, 
\]
and two dimensionless parameters $\gamma_{PN}$ and $\beta_{PN}$ which,
in the present theories turn out to be \cite{dam}
\begin{equation}
\gamma_{PN} -1= -2\frac{\alpha^2}{1+\alpha^2}\,,\,\,\,\, \beta_{PN} -1 = 
 \frac{\kappa \alpha^2}{(1+\alpha^2)^2}\,,
\label{postnewt}
\end{equation}
where $\kappa=\partial\alpha/\partial \varphi$.

A new constraint on the parameter $\gamma_{PN}$ has been obtained
recently using radio links with the Cassini spacecraft \cite{cassini},
\beq \gamma_{PN}-1 = (2.1\pm 2.3) \times 10^{-5}.
\label{casbound}
\eeq Present bounds on $\beta_{PN}-1$ are $O(10^{-4})$ and are less
restrictive for our choice of $\alpha(\vp)$, since $\kappa_0
\alpha^2(\vp_0) \simeq -\beta \alpha^3(\vp_0)$.

\begin{figure}[h]
\epsfxsize=3.2 in 
\epsfbox{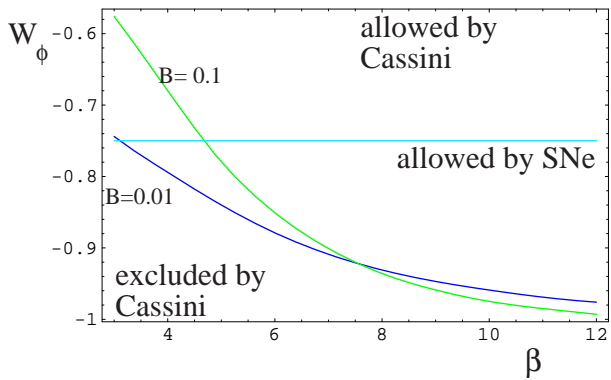}
\caption{The impact of the Cassini GR test. The regions below the
curves are excluded at $1 \sigma$ level. The SNIa bound on the DE
equation of state is also shown.}
\label{betaW}
\end{figure}

The bound from the Cassini spacecraft turns out to be quite strong
when used in connection with the one on the equation of state $w_\vp$
from SNe Ia. In Fig.~(\ref{betaW}) we show the excluded region in the
$\beta$-$w_\vp$ plane implied by Eq.~(\ref{casbound}). We see that an
equation of state $w_\vp < -0.78$, as implied by Sne Ia data at $95
\%$ c.l. \cite{SNe}, requires either a large value for $\beta$, or a
very small $B$. Since the last case corresponds to an expansion
history of the universe practically indistinguishable from GR, any
non-standard behavior induced by the ST theories in the past should be
likely accompanied by an equation of state different from $-1$ today.
If DE is a pure cosmological constant, then the bound from Cassini
implies $B<O(10^{-3})$ (making $A(\vp)$ practically indistinguishable
from one at least since BBN on), or unnaturally large values of
$\beta$.

\subsection{CMB power spectrum}
The impact of a cosmological constant or quintessence on the CMB power
spectrum has been extensively analyzed in refs. \cite{DEonCMB}. In the
context of ST theories, the problem has been studied in Refs.
\cite{Riaz,Sabino}. The main change with respect to a theory for DE
based on GR is due to a difference in the expansion rate, which
affects the angular scale of the anisotropies. The angle under which
the first peak is seen goes as 
\beq \theta_{\rm peak} \sim \pi/l_{\rm
peak} \sim v_s \, t_{\rm dec}\, z_{\rm dec}/ d(z_{\rm dec})\,, 
\eeq
where $l_{\rm peak}$ is the corresponding multipole, $v_s$ is the
sound speed, $t_{\rm dec}$ and $z_{\rm dec}$ are the time and redshift of
decoupling, and $d(z_{\rm dec})$ the distance to the last scattering
surface. The latter is given by 
\beq d(z_{\rm dec}) = \int_{1/(z_{\rm
dec}+1)}^1 \, \frac{d \tilde{a}}{\tilde{a} \tilde{H}}\,,
\label{dlast}
\eeq 
and is thus dominated by the behavior of $\tilde{H}$ close to the
upper limit of integration, where $\tilde{a} \tilde{H}$ is
smaller. For this reason, ST theories passing the GR tests ($A\to 1$
today, that is, $\tilde{H} \to \tilde{H}_{\rm GR}$) imply a small deviation of the
distance to the last scattering surface with respect to GR.

On the other hand, the decoupling time might be significantly more
perturbed. It is given by an expression analogous to Eq.~(\ref{dlast})
with the upper (lower) limit of integration replaced by $1/(z_{\rm
dec}+1)$ ($0$). As a result, since the universe expanded faster than
in GR at early times, we expect $t_{\rm dec}$ to be smaller, and the
peak to move towards higher multiples. We find 
\beq 
\frac{\Delta
l_{\rm peak}}{l_{\rm peak} }\simeq \frac{4}{3}\frac{A(\tilde{a}_{\rm
dec})-1}{A(\tilde{a}_{\rm dec})}\,, 
\eeq 
which is consistent with the
numerical findings of ref.~\cite{Riaz}.

Due to the well known degeneracy of the CMB spectrum with respect to
cosmological parameters, present measurements of the peak locations
\cite{CMB} do not translate straightforwardly into a bound on
$A(\tilde{a}_{\rm dec})$. For instance, it is found that a shift in
the peak multipole can be obtained also by varying the energy
densities according to \cite{huth-turner}
\begin{eqnarray}
 \frac{\Delta l_{\rm peak}}{l_{\rm peak}
} \simeq && -1.25 \frac{\Delta \Omega}{\Omega} \nonumber \\
&& -0.23 \frac{\Delta \Omega_M
h^2}{\Omega_M h^2} + 0.09 \frac{\Delta \Omega_b h^2}{\Omega_b h^2}
+ 0.089 \frac{\Delta \Omega_M}{\Omega_M}\,,
\end{eqnarray}
so that, in general, a full reanalysis of the CMB including the
new parameter $A(\tilde{a}_{\rm dec})$ would be required. However, in
the present case we find that, once the BBN bound on $A$ has been
imposed, the resulting values for $A(\tilde{a}_{\rm dec})$ are so
close to unity as to give shifts in the peak multiples smaller than
the experimental error. Thus, the CMB spectrum does not provide
significant bounds to the present scenario.

%
%
\section{Impact on WIMP relic abundance}
%

Having in mind all the bounds discussed in the previous Section, we can now go
on to compute the cosmological evolution of the scalar field and its 
impact on the DM relic abundance.

As a first step, we want to estimate if ST can have a sizeable effect on the
Jordan-frame Hubble parameter $\tilde{H}$ at the time of WIMP 
decoupling, without violating any of the avaliable cosmological observations.
We will consider the function $A(\varphi)$
of Eq.~(\ref{aphi}), imposing on the parameters $B$ and $\beta$
the phenomenological constraints already discussed.
We will then compute the ratio 
$\tilde{H}/\tilde{H}_{\rm GR}$ at the decoupling time of a typical WIMP of 
mass $m=200$~GeV. In this way we will be able to get an estimate of the effect 
before going into further detail.

The tightest bound is that coming from Eq.~(\ref{BBN}). It has an
impact on both $B$ in Eq.~(\ref{aphi}) and on the
initial conditions of $\vp$ at temperatures higher than the WIMP
freeze-out.  Indeed, since on the tracker solution the scalar field is
$\vp_0^{\rm tr} =O(1)$ today, it should have been $\ll 1$ at
nucleosynthesis, otherwise it would not have reached the attractor in
time \cite{tracker}. This implies $B\le O(0.1)$.

\begin{figure}[t!]
\epsfxsize=3.2 in 
\epsfbox{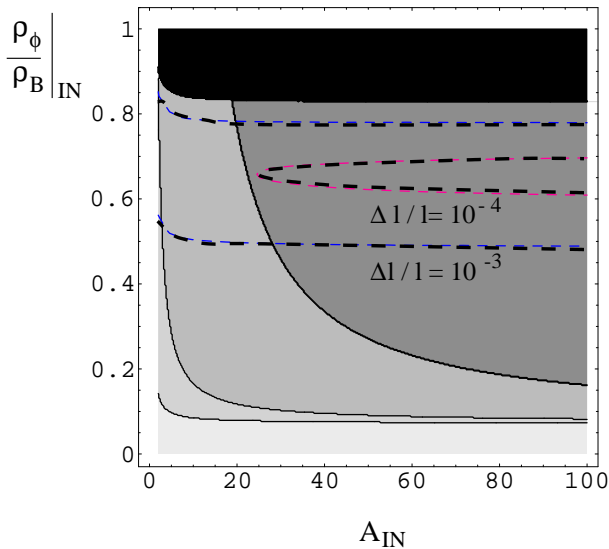}
\caption{The contours show the expansion rate enhancement
$\tilde{H}/\tilde{H}_{\rm GR}$ at $T=10$~GeV obtained in the ST model, as a
function of the initial values of the factor $A(\vp )$ and of the
ratio of the scalar to background energy density $\rho_\vp /
\rho_B$. We considered for the initial conditions a temperature of $T=
500$ GeV. 
The black area represents initial conditions which are excluded by nuclesynthesis.
The grey contours represent enhancements of $1 \div 10^2$, $10^2 \div 10^3$, 
$10^3 \div 10^4$, $10^4 \div 10^5$ from the lightest to the darkest. The
dashed lines show the shifts of the CMB doppler peaks obtained in the
ST model.}
\label{contorno}
\end{figure}

As already discussed, the equation for the dynamics of the scalar
field $\varphi$ is obtained by substituting the expression of
Eq.~(\ref{RHS}) in the RHS of Eq.~(\ref{eom}) and choosing a coupling
function $\alpha(\varphi)$ as defined in Eq.~(\ref{alphaphi}).  In the
sum of Eq.~(\ref{RHS}) only the terms corresponding to particles with
$m < T_c$ have been considered, i.e. particles lighter than the
critical temperature of the phase transition through which they
acquire a mass (see Ref. \cite{dam3b}). In particular we have taken
into account the top quark, the $Z^{0}$, the $W^{\pm}$, the bottom
quark, the tau quark, the charmed quark, the pions, the muon,
the electron and a WIMP particle of mass $m=200$~GeV.  
Numerical integration of the equation for
$\varphi$ has been carried out between approximately 500~GeV and today.
We have then computed, through Eq.(\ref{Htilde2}), the modified
expansion rate in the ST theory at a temperature $T=10$~GeV
corresponding to a typical time of WIMP decoupling and compared it to
the expansion rate of the standard case at the same temperature.

In Fig. \ref{contorno} we plot the ratio $\tilde{H}/\tilde{H}_{\rm GR}$ at
$T=10$~GeV as a function of the initial value of $A(\vp)$ and initial
ratio of the scalar to background energy density $\rho_\vp /
\rho_B$. We have restricted the possible initial conditions to those
regions of parameters values respecting the BBN bound of
Eq.~(\ref{BBN}).
We see that we have been able to produce an enhancement of the
expansion rate up to $O(10^5)$ at the time of WIMP decoupling. 
It is then worth studying in more detail what happens to the WIMP relic abundance.


Let us now consider the calculation of the relic abundance of a DM WIMP
with mass $m$ and annihilation cross-section $\langle\sigma_{\rm ann}
v\rangle$. As already mentioned, laboratory clocks and rods measure
the ``physical'' metric $\tilde{g}_{\mu\nu}$ and so the standard laws
of non-gravitational physics take their usual form in units of the
interval $d\tilde{s}^2$. As outlined in Ref.\cite{damour-pichon}, the
effect of the modified ST gravity will enter the computation of
particle physics processes (like the WIMP relic abundance) through the
``physical'' expansion rate $\tilde{H}$ defined in Eq.~(\ref{Htilde}).
We have therefore to implement the standard Boltzmann equation with
the modified physical Hubble parameter $\tilde{H}$:
\begin{equation}
\frac{dY}{dx} = -\frac{1}{x} \frac{s}{\tilde{H}} \langle\sigma_{\rm
ann} v\rangle (Y^2 - Y_{\rm eq}^2)
\label{eq:boltzmann}
\end{equation}
where $x=m/T$, $s=(2\pi^2/45)~h_\star(T)~T^3$ is the entropy density
and $Y=n/s$ is the WIMP density per comoving volume.

\begin{figure}[b] \centering
\vspace{-20pt}
\includegraphics[width=1.0\columnwidth]{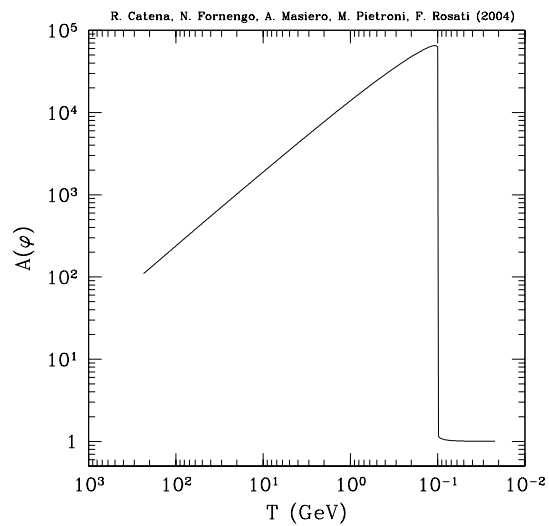}
\vspace{-20pt}
\caption{\label{fig:aphi} A typical behaviour of the function
$A(\varphi)$ defined in Eq.~(\ref{aphi}), calculated for parameters
$B=0.1$ and $\beta=8$.}
\end{figure}

We have considered values of $\tilde{H}$ wich respect all the bounds
discussed in Section IV. Specifically, we have considered the function
$A(\varphi)$ as given in Eq.~(\ref{aphi}) with parameters $B=0.1$ and
$\beta=8$. The function $A(\varphi)$ for this choice of parameters is
plotted in Fig.~\ref{fig:aphi}, which shows that $A(\varphi)$ is very
large at large temperatures, and then, at a temperature $T_\varphi$,
sharply drops to values close to 1 before nucleosynthesis sets in. A
parametrization of the behaviour of $A(\varphi)$ for $T>T_\varphi$,
that will be useful in the following discussion, is:
\begin{eqnarray}
A(\varphi) &\simeq& 2.19\cdot 10^{14} \left(\frac{T_0}{T}\right)^{0.82} \nonumber \\
&\simeq& 9.65\cdot 10^3 \left(\frac{\rm GeV}{m}\right)^{0.82}\, x^{0.82}
\label{eq:aphi}
\end{eqnarray}
where $T_0$ is the current temperature of the Universe.

We have numerically checked that, in the regime we are considering, a
good approximation to the physical Hubble parameter is given by:
\begin{equation}
\tilde{H} = A(\varphi) \tilde{H}_{\rm GR}
\end{equation}
The solution of the Boltzmann equation is therefore formally the same
as in the standard case, with the noticeable difference that now the
Hubble parameter gets an additional temperature dependence, given by
the function $A(\varphi)$. This can be translated in a change in the
effective number of degrees of freedom at temperature $T$:
\begin{equation}
g_\star(x)  \longrightarrow A^2(x) g_\star(x)
\end{equation}
An approximated solution of Eq.(\ref{eq:boltzmann}) can be cast
in a form analogous to the standard case:
\begin{equation}
\frac{1}{Y_0} = \frac{1}{Y_f} + \sqrt{\frac{\pi}{45 G}}~ m
\int_{x_f}^{\infty} dx\, \frac{A^{-1}(x)\,G(x)\langle\sigma_{\rm ann}
v\rangle}{x^2}
\label{eq:Y0}
\end{equation}
where $G(x)=h_\star(x)/g^{1/2}_\star(x)$ and $Y_0$ and $Y_f$ are the
WIMP abundances per comovin volume today and at freeze--out,
respectively.
\begin{figure}[t] \centering
\vspace{-20pt}
\includegraphics[width=1.0\columnwidth]{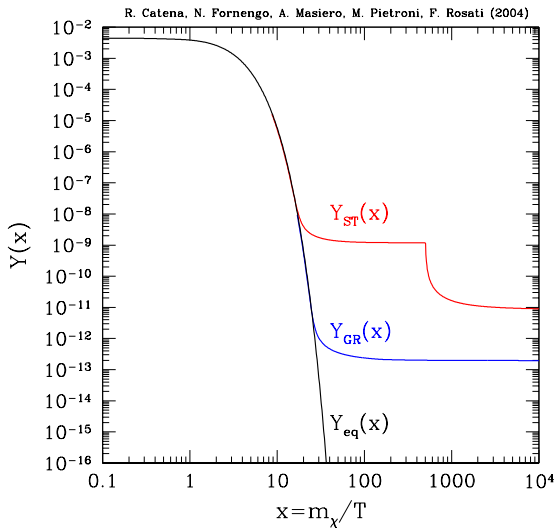}
\vspace{-20pt}
\caption{\label{fig:abundance} Numerical solution of the Boltzmann
equation Eq.~(\ref{eq:boltzmann}) in a ST cosmology for a toy--model
of a DM WIMP of mass $m=50$ GeV and constant annihilation
cross-section $\langle\sigma_{\rm ann} v\rangle = 1\times 10^{-7}$
GeV$^{-2}$. The temperature evolution of the WIMP abundance $Y(x)$
clearly shows that freeze--out is anticipated, since the expansion
rate of the Universe is largely enhanced by the presence of the scalar
field $\varphi$. At a value $x=m/T_\varphi$ a re--annihilation phase
occurs and $Y(x)$ drops to the present day value.}
\end{figure}
The freeze--out temperature is obtained by the
following implicit equation:
\begin{equation}
x_f = \ln \left[0.038\,M_P\,g\,m\, \frac{\langle\sigma_{\rm ann}
v\rangle_f\,x_f^{-1/2}}{A(x_f)g^{1/2}_\star(x_f)}\right]
\label{eq:xf}
\end{equation}
where $g$ is the internal number of degrees of freedom of our WIMP.
Clearly, when $A(x)\rightarrow 1$ we recover the standard case.
The relic abundance is then simply given by:
\begin{equation}
\Omega h^2 = \frac{m\,s_0\,Y_0}{\rho_{\rm crit}}
\end{equation}
where $s_0$ is the present entropy density and $\rho_{\rm crit}$
denotes the critical density.

A numerical solution of the Boltzmann equation Eq.~(\ref{eq:boltzmann})
in a ST cosmology with the function $A(x)$ given in
Fig.~\ref{fig:aphi} is shown in Fig.~\ref{fig:abundance} for a
toy--model of a DM WIMP of mass $m=50$ GeV and constant annihilation
cross-section $\langle\sigma_{\rm ann} v\rangle = 1\times 10^{-7}$
GeV$^{-2}$. The temperature evolution of the WIMP abundance $Y(x)$
clearly shows that freeze--out is anticipated, since the expansion
rate of the Universe is largely enhanced by the presence of the scalar
field $\varphi$. This effect is expected. However, we note that a
peculiar effect emerges: when the ST theory approached GR (a fact
which is parametrized by $A(\vp) \rightarrow 1$ at a temperature
$T_\varphi$, which in our model is 0.1 GeV), $\tilde H$ rapidly drops
below the interaction rate $\Gamma$ establishing a short period during
which the already frozen WIMPs are still abundant enough to start a
sizeable re--annihilation. This post-freeze--out ``re--annihilation
phase'' has the effect of reducing the WIMP abundance, which
nevertheless remains much larger than in the standard case. For the
specific case shown in Fig.~\ref{fig:abundance} the WIMP relic
abundance is $\Omega h^2=0.0027$ for GR, while for a ST cosmology
becomes $\Omega h^2=0.12$, with an increase of a factor of 44.

\begin{figure}[t] \centering
\vspace{-20pt}
\includegraphics[width=1.0\columnwidth]{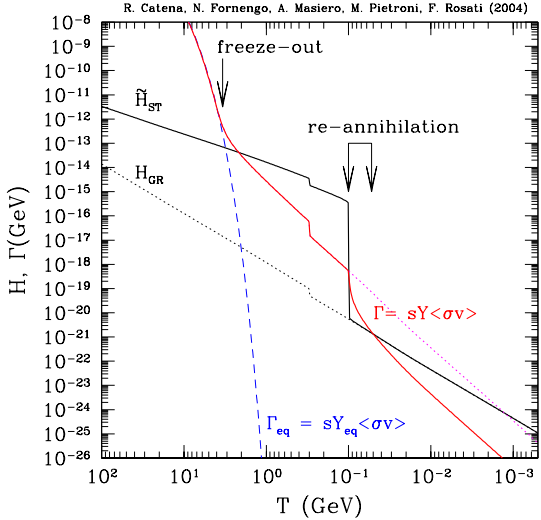}
\vspace{-20pt}
\caption{\label{fig:rates} The Expansion rate of the Universe $\tilde
H$ and the WIMP interaction rate $\Gamma=Y\, s\, \langle\sigma_{\rm
ann} v\rangle $ are plotted as a function of the temperature. The
re-annihilation effect discussed in the text is outlined. The small
drop in the rates at $T=300$ MeV is due to the quark--hadron phase
transition.}
\end{figure}

The phenomenon of re--annihilation can be conveniently discussed in
terms of the relation between the expansion rate of the Universe
$\tilde H$ and the WIMP interaction rate $\Gamma=Y\, s\,
\langle\sigma_{\rm ann} v\rangle $. A numerical calculation of these
two quantities is plotted in Fig.~\ref{fig:rates} as a function of the
temperature. The departure from equilibrium occurs earlier than in the
GR case, because $\tilde H \gg \tilde{H}_{\rm GR}$. When decoupling is
completed, the particles evolve with an approximately constant $Y=Y_f$
and $\Gamma \sim T^3$, while the Hubble rate evolves as $\tilde{H}
\sim A(x)\,\tilde{\rho}^{1/2}\,\sim T^{1.2}$, i.e. slower than in the
standard case (we have used here the approximate $A(x)$ behavious of
Eq.(\ref{eq:aphi})).

At the transition temperature
$T_\varphi$ the Hubble rate drops to its standard value $H_{\rm GR}$
and becomes smaller than the interaction rate: in this case the
decoupled WIMPs start to annihilate again, for a short period. After
this re--annihilation phase, the particles continue to evolve with an
approximately constant abundace $Y<Y_f$ and $\Gamma$ recovers the
behaviour $T^3$, while $\tilde{H}_{\rm GR} \sim \tilde{\rho}^{1/2}\sim
T^2$ as usual.

\begin{figure}[t] \centering
\vspace{-20pt}
\includegraphics[width=1.0\columnwidth]{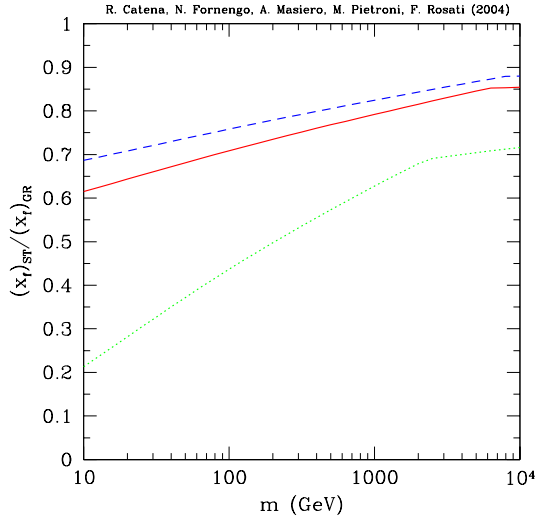}
\vspace{-20pt}
\caption{\label{fig:freeze} The ratio between the freeze--out values
of $x_f=m/T_f$ in ST cosmology and in GR as a function of the WIMP
mass.  The dashed, solid and dotted lines refer to $\langle\sigma_{\rm
ann} v\rangle=10^{-4}$ GeV$^{-2}$, $10^{-7}$ GeV$^{-2}$ and $10^{-14}$
GeV$^{-2}$, respectively.}
\end{figure}

We notice that a re--annihilation phase does not occur in the case of
kination, i.e. in the case the energy density of the Universe is
dominated by the kinetic term of a scalar field
\cite{salatirosati}. In this case the evolution of the expansion rate
is $\tilde{H}_{\rm kin} \sim T^3$ during kination, and than evolves
smootly into the standard behaviour $\tilde{H}_{\rm GR} \sim T^2$.
During kination both $\tilde{H}$ and $\Gamma$ have the same
$T$--dependence and closely follow each other, until kination ends and
the standard behaviour is recovered. Re--annihilation is possibile if
the phase during which the expansion rate has the transition toward
its standard GR behaviour is faster than the post-freeze--out
evolution of the interaction rate, i.e. faster than $T^3$.

The change in the freeze--out temperature is shown in
Fig.~\ref{fig:freeze} where we show the ratio between the freeze--out
values of $x_f=m/T_f$ in ST cosmology and in GR. The freeze--out
temperature is anticipated about a factor of 2, with a dependence also
on the annihilation cross section, as is clear from Eq.~(\ref{eq:xf}):
for very low values of $\langle\sigma_{\rm ann} v\rangle$ the
freeze-out temperature may be anticipated up to a factor of 5. For
these low cross sections the relic abundance is anyway largely
overabundant: we can therefore quantify the reduction in $x_f$ in a
factor which ranges between 10\% 
and 40\% 
for WIMPs which can provide
abundances in the cosmologically acceptable range.

\begin{figure}[t] \centering
\vspace{-20pt}
\includegraphics[width=1.0\columnwidth]{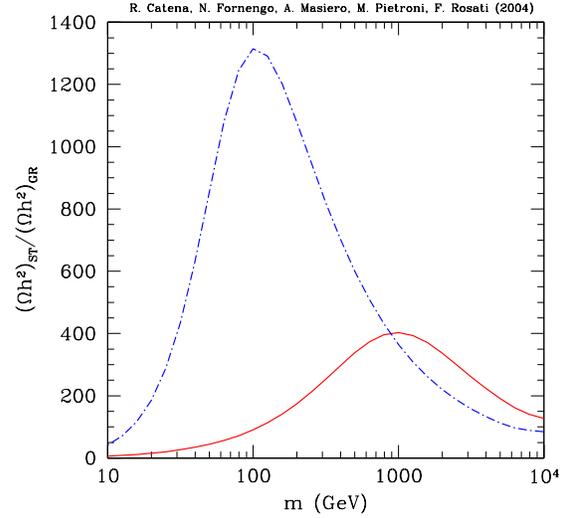}
\vspace{-20pt}
\caption{\label{fig:ratio} Increase in the WIMP relic abundance in ST
cosmology with respect to the GR case.  The solid curve refers to an
annihilation cross section constant in temperature, i.e.
$\langle\sigma_{\rm ann} v\rangle = a = 10^{-7}$ GeV$^{-2}$, while the
dashed line stands for an annihilation cross section which evolves
with temperature as $\langle\sigma_{\rm ann} v\rangle = b/x = 10^{-7}$ GeV$^{-2}/x$.}
\end{figure}

The amount of increase in the relic abundance which is present in ST
cosmology is shown in Fig.~\ref{fig:ratio}. The solid curve refers to
an annihilation cross section constant in temperature, i.e.
$\langle\sigma_{\rm ann} v\rangle = a$, while the dashed line stands
for an annihilation cross section which evolves with temperature as:
$\langle\sigma_{\rm ann} v\rangle = b/x$ (these two cases correspond
to the two limiting situations of the usual non--relativistic
expansion of the thermally averaged annihilation cross section:
$\langle\sigma_{\rm ann} v\rangle = a + b/x$). In the case of
$s$--wave annihilation the increase in relic abundance ranges from a
factor of 10 up to a factor of 400. For a pure $b/x$ dependence, the
enhancement can be as large as 3 orders of magnitude.

\begin{figure}[t] \centering
\vspace{-20pt}
\includegraphics[width=1.0\columnwidth]{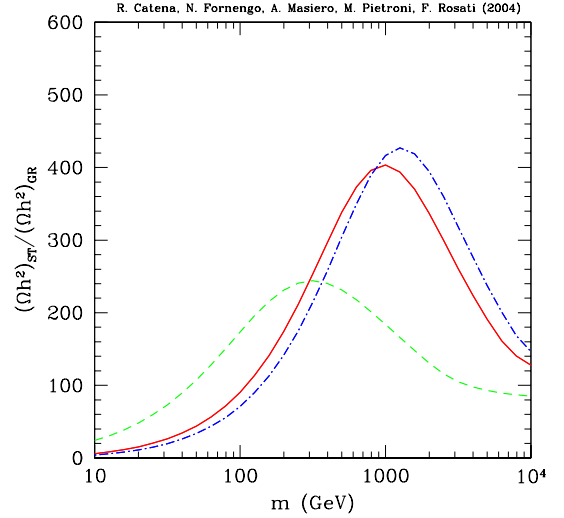}
\vspace{-20pt}
\caption{\label{fig:ratioa} Increase in the WIMP relic abundance with
an annihilation cross section $\langle\sigma_{\rm ann} v\rangle = a$,
for different values of $a$. The dot--dashed, solid and dashed lines
correspond to $a=10^{-4}$ GeV$^{-2}$, $10^{-7}$ GeV$^{-2}$ and
$10^{-14}$ GeV$^{-2}$, respectively.}
\end{figure}

The behaviours shown in Fig.~\ref{fig:ratio}, which have been obtained
by a numerical integration of the Boltzmann equation
Eq.~(\ref{eq:boltzmann}), can be understood by employing the
approximate analytical solution (\ref{eq:Y0}). In the case of
$\langle\sigma_{\rm ann} v\rangle = a$, Eq.~(\ref{eq:Y0}) gives:
\begin{equation}
\frac{1}{Y_0} = {\cal G}\,m\,G(x^{\rm GR}_f)\frac{a}{x^{\rm GR}_f}
\label{eq:Ystd}
\end{equation}
in the standard GR case, and
\begin{eqnarray}
\frac{1}{Y_0} &=& 
 {\cal G}\,m \left[\frac{G(x_f^{ST})}{\bar
A} \frac{a}{1.82}\left(\frac{1}{(x_f^{ST})^{1.82}}-\frac{1}{(x_\varphi)^{1.82}}\right) \right] \nonumber \\
&+& {\cal G}\,m \left[G(x_\varphi)\frac{a}{x_\varphi}\right]
\label{eq:Yphi}
\end{eqnarray}
in our ST model where the $A(x)$ function is given in
Eq.~(\ref{eq:aphi}) for $T>T_\varphi$ and $A(x)=1$ otherwise (${\bar A}
= 9.65\cdot 10^3 ({\rm GeV}/m)^{0.82}$).  For the sake of simplicity,
in both solutions we have dropped the term $1/Y_f$ which adds a small
correction, not relevant for the present approximate discussion. In
both equations ${\cal G} = \sqrt{\pi/(45\,G)}$. The ratio $R$ of the
relic abundances is:
\begin{eqnarray}
R\equiv\frac{(\Omega h^2)_{\rm ST}}{(\Omega h^2)_{\rm GR}} \simeq
\frac{1.82\,\bar{A}\,x_\varphi\,x_f^{0.82}}{x_\varphi+1.82\,\bar{A}\,r_G\,x_f^{1.82}}
\label{eq:ratio0}
\end{eqnarray}
where we have approximated $x^{\rm GR}_f\simeq x^{ST}_f$ and we have
defined $r_G=G(x_\varphi)/G(x_f)$. By making explicit the mass
dependencies we obtain:
\begin{eqnarray}
R \simeq \frac{A_R\,m_{\rm GeV}}{B_R + m_{\rm Gev}^{1.82}}
\label{eq:ratio}
\end{eqnarray}
where the mass is expresses in GeV, $A_R  \simeq 1.76\cdot
10^4\,x_f^{0.82}\simeq 2.05 \cdot 10^{5}$, $B_R=1.76\cdot
10^4\,T_\varphi\,r_G\,x_f^{1.82}\simeq 2.05 \cdot 10^{5}$, and the
numerical values have been obtained for $x_f\simeq 20$ and $r_G\simeq
0.5$ (since in our case $T_\varphi$ is smaller than the quark--hadron
phase transition which we have set at $T_{\rm QCD} = 300$ MeV). The
analytic approximation of Eq.~(\ref{eq:ratio}) helps to explain the
behaviour shown by the solid curve in Fig.~\ref{fig:ratio}, which has
been obtained by numerical calculations which employ the exact form of
the function $A(\varphi)$.  From Eq.~(\ref{eq:ratio}) we can in fact
derive that, for low masses, the ratio $R$ has the behaviour:
\begin{equation}
R \simeq \frac{m}{r_G\,T_\varphi}\,\frac{1}{x_f} = \frac{1}{r_G}\,\frac{T_f}{T_\varphi}
\end{equation}
which shows that in this mass regime $R$ grows almost linearly with
the WIMP mass $m$, and it is larger for lower values of $T_\varphi$. If
we accept $T_\varphi$ as low as the BBN scale, we can obtain a further
increase in the relic abundance of a factor 100 on the top of the one
showed in Fig.~\ref{fig:ratioa} for low values of $m$.
When the WIMP mass is very large, the ratio $R$ behaves as:
\begin{equation}
R \simeq \frac{1.76\cdot 10^4 \, x_f^{0.82}}{m_{\rm GeV}^{0.82}}
\end{equation}
with a slight drop with the mass. The position of the maximum and the
maximal value of $R$ are given by:
\begin{equation}
m^{\rm max}_{\rm GeV} \simeq (2.15\cdot 10^4\,r_G\,T_\varphi)^{0.56}\,\,x_f
\end{equation}
and:
\begin{equation}
R_{\rm max} \simeq \frac{108}{(r_G T_\varphi)^{0.45}}
\end{equation}

These expressions show that the maximal effect is also obtained for
the lowest values of $T_\varphi$; in this case the position of $R_{\rm
max}$ is shifted toward lower masses. For $T_\varphi$ at the BBN scale,
the maximal increase in the WIMP relic abundance is of the order of
3000, instead of about 400 obtained for $T_\varphi=0.1$ GeV and shown
in Fig.~\ref{fig:ratioa}.

An interesting property shown by Eq.~(\ref{eq:ratio}) is that $R$ does
not depend explicitely on the annihilation cross section
$\langle\sigma_{\rm ann} v\rangle = a$, which drops out in the
ratio. An implicit dependence on the cross section is present through
$x_f$, as can be seen in Eq.~(\ref{eq:xf}). This dependence however is
only logarithmic and does not spoil the general behaviour of $R$ shown
in Fig.~\ref{fig:ratio}. This is shown in Fig.~\ref{fig:ratioa}: the
largest difference occurs for very low annihilation cross sections,
for which the deviation of $x_f$ is larger. However, for cross
sections of interest, i.e. cross section which provide relic
abundances below the cosmologically acceptable upper bound, the values
of $R$ are stable to a relatively good extent.

\begin{figure}[t] \centering
\vspace{-20pt}
\includegraphics[width=1.0\columnwidth]{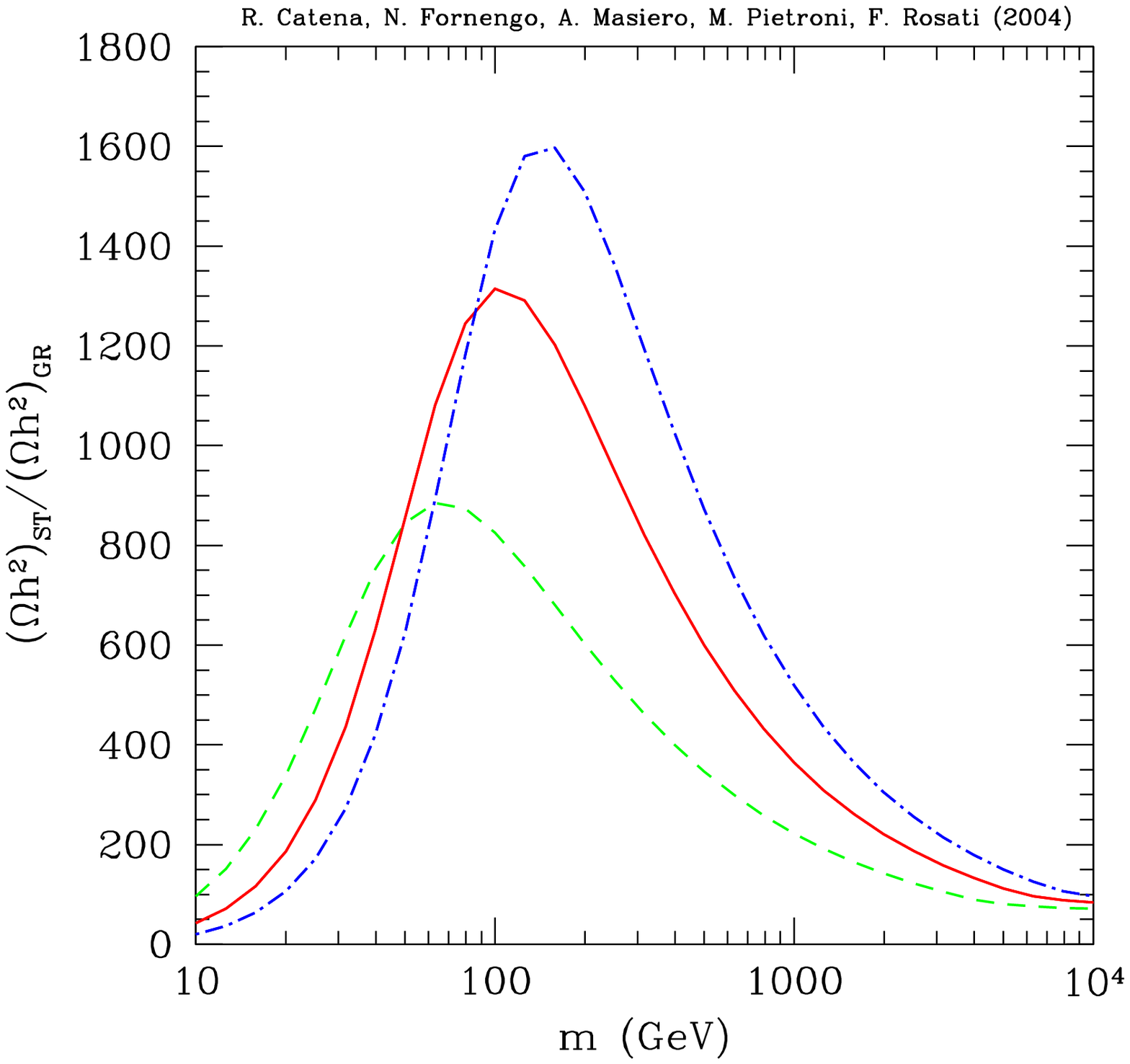}
\vspace{-20pt}
\caption{\label{fig:ratiob} Increase in the WIMP relic abundance with
an annihilation cross section $\langle\sigma_{\rm ann} v\rangle =
b/x$, for different values of $b$.  The dot--dashed, solid and dashed
lines correspond to $b=10^{-4}$ GeV$^{-2}$ $10^{-7}$ GeV$^{-2}$ and
$10^{-10}$ GeV$^{-2}$, respectively.}
\end{figure}

\begin{figure}[t] \centering
\vspace{-20pt}
\includegraphics[width=1.0\columnwidth]{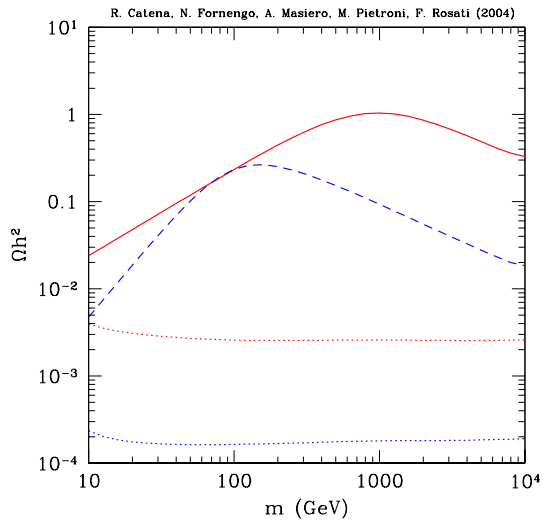}
\vspace{-20pt}
\caption{\label{fig:omega} The relic abundance in a ST theory as a
function of the WIMP mass in the case of $\langle\sigma_{\rm ann}
v\rangle \equiv a = 1\cdot 10^{-7}$ GeV (solid line) and
$\langle\sigma_{\rm ann} v\rangle \equiv b/x= 1\cdot 10^{-4}~{\rm
GeV}/x$ (dashed line).  The upper (lower) dotted lines corresponds to
the GR case for $\langle\sigma_{\rm ann} v\rangle \equiv a = 1\cdot
10^{-7}$ GeV and $\langle\sigma_{\rm ann} v\rangle \equiv b/x= 1\cdot
10^{-4}~{\rm GeV}/x$, respectively.}
\end{figure}

A similar analysis holds in the case of $\langle\sigma_{\rm ann}
v\rangle = b/x$. However, in this case the dependence of $R$ with
$x_f$ is somehow stronger (as obtained from the integration in
Eq.~(\ref{eq:Y0})), and the effect of changing $\langle\sigma_{\rm ann}
v\rangle$ is slightely larger. This effect can be seen in
Fig.~\ref{fig:ratiob}, where $R$ is shown for different values of the
parameter $b$.  Notice that larger cross sections, which in the
standard case provide lower values for the relic abundance, are the
ones which get more enhanced in ST cosmology.

Finally, as an example we show in Fig.~\ref{fig:omega} the relic
abundance as a function of the WIMP mass in the case of
$\langle\sigma_{\rm ann} v\rangle \equiv a = 1\cdot 10^{-7}$ GeV and
$\langle\sigma_{\rm ann} v\rangle \equiv b/x= 1\cdot 10^{-4}~{\rm
GeV}/x$.  We see that, in this case, the relic abundance can be at the
level required to explain the CDM content of the Universe
($\Omega_{\rm CDM} h^2 = 0.095 \div 0.13$ \cite{wmap}) for a ST
theory, while it is underabundant in the standard case. The models
shown in Fig.~\ref{fig:omega} represent a case in which we can explain
at the same time both the DM and DE contents of the Universe, and the
interplay of the two component is crucial in determinig the right
abundances of both DM and DE.

An analysis of specific particle candidates of DM, in particular in
supersymmetric models, will be examined elsewhere \cite{inprogress}.

%
%
\section{Conclusions}
%
The idea of exploiting primordial (ultralight) scalars in
order to shed some light on a dynamical interpretation of DE is by now
a widespread research topic in the literature. In this paper we follow
the promising proposal of considering the quintessence scalar as
embedded in a scalar--tensor theory of gravity.  This approach is
at variance with the usual interpretation of quintessence as a new
light scalar whose interactions with matter are subject to the tight
phenomenological constraints on the equivalence principle violation
and time-variation of the fundamental coupling constants. Identifying
the quintessence field with the scalar component of a ST theory, instead, 
does not pose any threat on the equivalence principle, since
by construction matter has a purely metric coupling with gravity. 
 
We focus on quintessence ST models which possess a double ``attraction mechanism'', 
one to GR and
the other ensuring $\rho_{DE}$ to follow a tracking solution. These two
simultaneous mechanisms act as a ``protection'' for the theory to prevent
its fall into immediate troubles (for instance, large
departures from GR predictions). 
Nevertheless, we still obtain important phnomenological signatures which might
disentangle this theory from GR or other alternative proposals. 
The tests of our ST scenario divide  into two classes: deviations from GR
and departures from standard cosmology, in particular concerning the
expansion rate of the Universe. 

The latter effect has a big impact on the most distant epoch of
the Universe for which we have ``direct'' information,
i.e. nucleosynthesis. However, we pointed out that we can further
extend the implications of a non-standard $\tilde{H}$ in the early
Universe to times prior to nucleosynthesis.  Sticking to the
standard WIMP picture of DM, one of the most relevant events before
BBN is the WIMP decoupling which is expected to have occurred 
at a temperature of a few GeVs. 
Our work shows that,
despite the severe ``filters'' on ST quintessence models which are
provided by BBN, solar system tests of gravity and, to a lesser
degree, by CMB, it is still possible to find remarkable enhancements
on the expansion rate of the Universe at  WIMP freeze-out, 
yielding to relic WIMP
abundances which can vary up to a few orders of magnitude with respect
to the standard case.

In this paper we pointed out  some general features of 
the new ``WIMP story''  around its decoupling temperature in the
presence of ST quintessence. In particular, we noticed that some
unexpected effect can take place, such as a short phase of WIMP ``re-annihilation''    
when ST approaches GR. 
Needless to say, such potentially  
(very) large deviations entail new prospects on the WIMP characterization
both for the choice of the CDM candidates and  for their direct and indirect
detection probes. A thorough reconsideration of the ``traditional''
WIMP identified with the lightest neutralino in SUSY extensions of the SM
as well as the identification of other potentially viable CDM candidates
in the ST context  is presently under way \cite{inprogress}.

%
%
\section{Aknowledgments}
%
We acknowledge Research Grants funded by the Italian Ministero
dell'Istruzione, dell'Universit\`a e della Ricerca (MIUR) and by the
Istituto Nazionale di Fisica Nucleare (INFN) within the {\sl
Astroparticle Physics Project}. FR is partially supported by the
University of Padova fund for young researchers, research project
n. $CPDG037114$.


\begin{thebibliography}{99}

\bibitem{rev-cosmo}
W.~L.~Freedman and M.~S.~Turner,
Rev.\ Mod.\ Phys.\  {\bf 75} (2003) 1433.

\bibitem{quint-review}
For recent reviews, see for example: 
S.~M.~Carroll, Living Rev.\ Rel.\  {\bf 4} (2001) 1;
P.~J.~E.~Peebles and B.~Ratra,
Rev.\ Mod.\ Phys.\  {\bf 75} (2003) 559;
T.~Padmanabhan, Phys.\ Rept.\  {\bf 380} (2003) 235.

\bibitem{carroll1}
S.~M.~Carroll,
Phys.\ Rev.\ Lett.\  {\bf 81} (1998) 3067.
\bibitem{mpr}
A.~Masiero, M.~Pietroni and F.~Rosati,
Phys.\ Rev.\ D {\bf 61} (2000) 023504.
\bibitem{dpv} T.~Damour, F.~Piazza and G.~Veneziano, Phys.\ Rev.\ D {\bf 66} (2002) 046007;
T.~Damour, F.~Piazza and G.~Veneziano, Phys.\ Rev.\ Lett.\  {\bf 89} (2002) 081601

\bibitem{dam3a} T.~Damour and K.~Nordtvedt, Phys. Rev. {\bf D48}, 3436
(1993).

\bibitem{dam3b} T.~Damour and A.M.~Polyakov, Nucl.\ Phys. {\bf B423},
532 (1994).

\bibitem{bd}  P.~Jordan, {\em Schwerkaft und Weltall} (Vieweg, Braunschweig,
1955); M. Fierz, Helv. Phys. Acta {\bf 29}, 128 (1956); C.~Brans and
R.H.~Dicke, Phys. Rev. {\bf 124}, 925 (1961).

\bibitem{max} 
N.~Bartolo and M.~Pietroni, Phys.\ Rev.\ D {\bf 61} (2000) 023518

\bibitem{tracker}
A.~R.~Liddle and R.~J.~Scherrer,
Phys.\ Rev.\ D {\bf 59} (1999) 023509;
I.~Zlatev, L.~M.~Wang and P.~J.~Steinhardt,
Phys.\ Rev.\ Lett.\  {\bf 82} (1999) 896;
P.~J.~Steinhardt, L.~M.~Wang and I.~Zlatev,
Phys.\ Rev.\ D {\bf 59} (1999) 123504;
B.~Ratra and P.~J.~E.~Peebles,
Phys.\ Rev.\ D {\bf 37} (1988) 3406;
P.~J.~E.~Peebles and B.~Ratra,
Astrophys.\ J.\  {\bf 325} (1988) L17.

\bibitem{joyce}
M.~Joyce and T.~Prokopec, JHEP {\bf 0010} (2000) 030;
M.~Joyce, Phys.\ Rev.\ D {\bf 55} (1997) 1875.

\bibitem{damour-pichon} T.~Damour and B.~Pichon, Phys.\ Rev.\ D {\bf 59} (1999) 123502.

\bibitem{santiago}
D.I.~Santiago, D.~Kalligas and R.V.~Wagoner, 
Phys. Rev. {\bf D58} (1998) 124005.

\bibitem{carroll2}
S.~M.~Carroll and M.~Kaplinghat,
Phys.\ Rev.\ D {\bf 65} (2002) 063507.

\bibitem{salatirosati} P.~Salati,
Phys.\ Lett.\ B {\bf 571}, 121 (2003);
F.~Rosati,
Phys.\ Lett.\ B {\bf 570}, 5 (2003).

\bibitem{ullio} S.~Profumo and P.~Ullio,
JCAP {\bf 0311} (2003) 006

\bibitem{comelli}
D.~Comelli, M.~Pietroni and A.~Riotto,
Phys.\ Lett.\ B {\bf 571} (2003) 115;
M.~Kamionkowski and M.~S.~Turner, Phys.\ Rev.\ D {\bf 42}, 3310 (1990); 
J.~D.~Barrow, Nucl.\ Phys.\ B {\bf 208} (1982) 501. 

\bibitem{Riaz} A.~Riazuelo and J.~P.~Uzan,
Phys.\ Rev.\ D {\bf 66}, 023525 (2002).

\bibitem{CMB} P.~de Bernardis {\it et al.},
Astrophys.\ J.\  {\bf 564}, 559 (2002); R.~Stompor {\it et al.},
Astrophys.\ J.\  {\bf 561}, L7 (2001); 
C.~L.~Bennett {\it et al.},
Astrophys.\ J.\ Suppl.\  {\bf 148}, 1 (2003).

\bibitem{cassini} B.~Bertotti, L.~Iess and P.~Tortora, Nature {\bf 425}, 374 (2003)

\bibitem{SNe} 
A.~G.~Riess {\it et al.}  [Supernova Search Team Collaboration],
Astron.\ J.\  {\bf 116}, 1009 (1998);
S.~Perlmutter {\it et al.}  [Supernova Cosmology Project Collaboration],
Astrophys.\ J.\  {\bf 517}, 565 (1999);
J.~L.~Tonry {\it et al.},
Astrophys.\ J.\  {\bf 594}, 1 (2003);
B.~J.~Barris {\it et al.},
Astrophys.~J. {\bf 602}, 571 (2004);
A.~G.~Riess {\it et al.},
Astrophys.~J. {\bf 607}, 665 (2004).

\bibitem{inprogress}
R.~Catena, N.~Fornengo, A.~Masiero, M.~Pietroni and F.~Rosati, in progress

\bibitem{dam}  T.~Damour,  gr-qc/9606079, lectures given at Les Houches
1992, SUSY95 and Corf\'{u} 1995.;
G.~Esposito-Farese and D.~Polarski, Phys.\ Rev.\ D {\bf 63}, 063504 (2001); 
B.~Boisseau, G.~Esposito-Farese, D.~Polarski and A.~A.~Starobinsky, 
Phys.\ Rev.\ Lett.\  {\bf 85}, 2236 (2000). 

\bibitem{sabino2}
S.~Matarrese, C.~Baccigalupi and F.~Perrotta,
astro-ph/0403480.

\bibitem{nubound} E.~Lisi, S.~Sarkar and F.~L.~Villante,
Phys.\ Rev.\ D {\bf 59} (1999) 123520; K.~A.~Olive and D.~Thomas,
Astropart.\ Phys.\  {\bf 11}, 403 (1999).

\bibitem{DEonCMB}
D.~Huterer and M.~S.~Turner,
Phys.\ Rev.\ D {\bf 64}, 123527 (2001);
S.~Hannestad and E.~Mortsell,
Phys.\ Rev.\ D {\bf 66}, 063508 (2002);
A.~Balbi, C.~Baccigalupi, F.~Perrotta, S.~Matarrese and N.~Vittorio,
Astrophys.\ J.\  {\bf 588}, L5 (2003).

\bibitem{Sabino} 
X.~l.~Chen and M.~Kamionkowski,
Phys.\ Rev.\ D {\bf 60}, 104036 (1999); 
F.~Perrotta, C.~Baccigalupi and S.~Matarrese,
Phys.\ Rev.\ D {\bf 61}, 023507 (2000);
R.~Nagata, T.~Chiba and N.~Sugiyama, Phys.\ Rev.\ D {\bf 69}, 083512 (2004); 
R.~Nagata, T.~Chiba and N.~Sugiyama, Phys.\ Rev.\ D {\bf 66}, 103510 (2002). 

\bibitem{huth-turner}
D.~Huterer and M.~S.~Turner,
Phys.\ Rev.\ D {\bf 64} (2001) 123527

\bibitem{wmap} 
D.N. Spergel {\it et al.}, Astrophys. J. Suppl. {\bf 148}, 175 (2003).


\end{thebibliography}
\end{document}